# Towards Efficient Image Deblurring for Edge Deployment

Srinivas Soumitri Miriyala*
*On-Device AI*
*Samsung Research Institute*
Bangalore, India
srinivas.m1@samsung.com

Sowmya Lahari Vajrala*
*On-Device AI*
*Samsung Research Institute*
Bangalore, India
v.lahari@samsung.com

Rama Sravanth Kodavanti
*On-Device AI*
*Samsung Research Institute*
Bangalore, India
k.sravanth@samsung.com

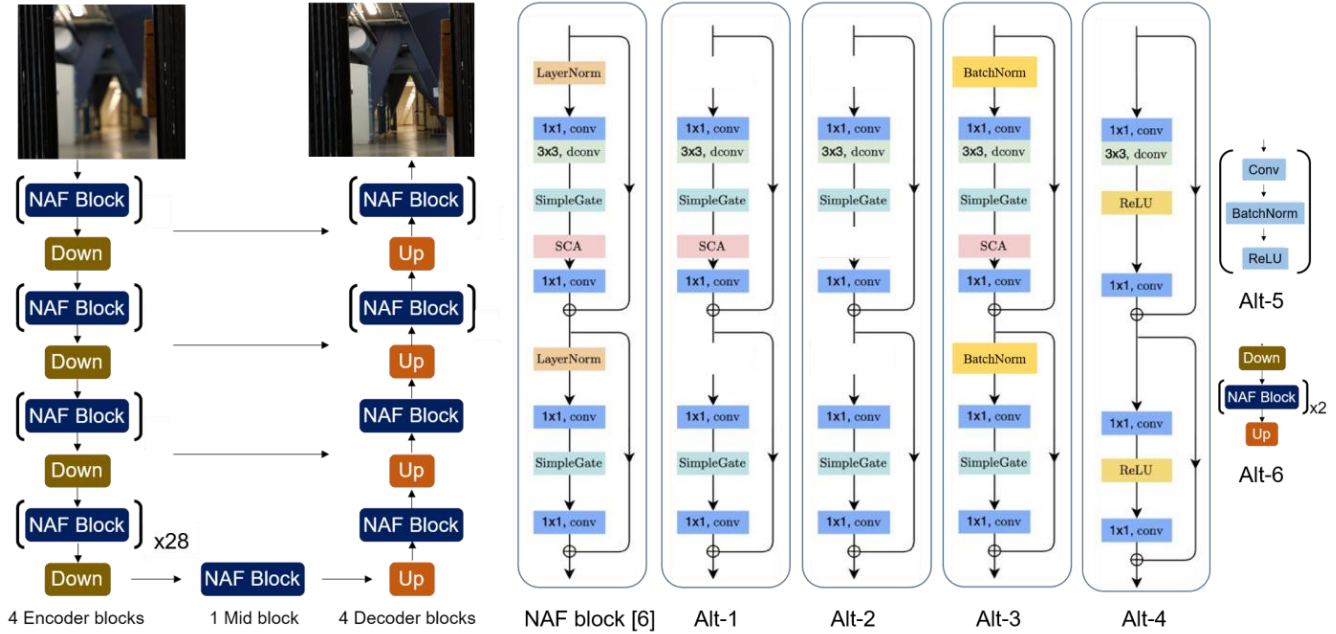

Figure 1: The base NAFNet [3] for Image De-blurring, NAF block and its 6 hardware-friendly alternatives proposed in this work.

*Abstract* - Image deblurring is a critical stage in mobile image signal processing pipelines, where the ability to restore fine structures and textures must be balanced with real-time constraints on edge devices. While recent deep networks such as transformers and activation-free architectures achieve state-of-the-art (SOTA) accuracy, their efficiency is typically measured in FLOPs or parameters, which do not correlate with latency on embedded hardware. We propose a hardware-aware adaptation framework that restructures existing models through sensitivity-guided block substitution, surrogate distillation, and training-free multi-objective search driven by device profiling. Applied to the 36-block NAFNet baseline, the optimized variants achieve up to 55% reduction in GMACs compared to recent transformer-based SOTA while maintaining competitive accuracy. Most importantly, on-device deployment yields a 1.25× latency improvement over the baseline. Experiments on motion deblurring (GoPro), defocus deblurring (DPDD), and auxiliary benchmarks (RealBlur-J/R, HIDE) demonstrate the generality of the approach, while comparisons with prior efficient baselines confirm its accuracy-efficiency trade-off. These results establish feedback-driven adaptation as a principled strategy for bridging the gap between algorithmic design and deployment-ready deblurring models.

## I. INTRODUCTION

Image deblurring is a central problem in computational photography, aiming to restore sharp images from inputs degraded by motion or defocus blur. It has also become a critical stage in mobile image signal processing pipelines, where the ability to reconstruct fine structures and textures directly impacts perceived visual quality. Over the past decade, deep learning has advanced this field substantially, with convolutional, transformer-based, and even diffusion-based models pushing state-of-the-art (SOTA) restoration accuracy. The key challenge is no longer how to build accurate deblurring networks, but how to adapt them for real-time deployment on opaque, resource-constrained hardware accelerators.

Early emphasis on efficiency came from Chiang et al. [1], who studied quality - latency trade-offs in deploying deblurring networks on mobile devices, highlighting quantization and operator selection for practical performance. Lightweight CNN approaches soon followed: HINet [2] introduced Half Instance Normalization to balance efficiency and accuracy, while NAFNet [3] simplified architecture by removing nonlinear activations and relying on lightweight gating. Transformer-based designs such as Uformer [4] and Restormer [5] leveraged windowed attention and channel attention, respectively, to extend receptive fields while maintaining efficiency. Kernel-driven methods like KBNet [14] further reduced cost by learning kernel bases with multi-axis fusion.

These networks are often considered "efficient" when measured in FLOPs or parameter counts, but such metrics rarely capture the realities of deployment on edge devices. In practice, models optimized under these proxies may achieve strong benchmark scores yet fail to deliver real-time performance. For example, models like NAFNet, efficient in theory, still suffer latency from channel-attention modules on specialized embedded hardware accelerators. This gap motivates the need for frameworks that adapt existing high-performing architectures into truly hardware-efficient variants.

We address this challenge by proposing a hardware-aware optimization approach that leverages direct device feedback to restructure models without discarding their robustness. To demonstrate generality, we apply the framework to deblurring with NAFNet as baseline and studied the following tasks comprehensively.

a) Motion de-blurring on the GoPro dataset [8], with ~17 million parameter NAFNet model as base (see Fig. 1),

b) Testing the generalizability of model for motion de-blurring Real Blur-J, R [10] and HIDE [11].

c) Defocus de-blurring of DPDD dataset [9] to check the scope of adapting to multiple use-cases.

d) Comparison with prior-arts SOTA in accuracy and other resource-efficient baselines.

e) Deployment & profiling on the Neural Processing Unit (NPU) of Samsung Galaxy S24 (GS24) Ultra.

Table 1: Saliency scores of all the blocks in the NAFNet GoPro

| Block | Grad norm | SNIP | GraSP | Fisher (e-06) | Plain (e-03) | Synflow |
|---|---|---|---|---|---|---|
| Enc0 | 0.0118 | 0.04 | 0.0005 | 1.96 | -6.95 | 2.0e+18 |
| Enc1 | 0.0050 | 0.03 | -0.0001 | 2.58 | -3.72 | 2.3e+16 |
| Enc2 | 0.0037 | 0.05 | 0.0006 | 5.19 | 2.24e | 1.6e+15 |
| Enc3 | 0.0644 | 1.6 | -0.0017 | 6.23 | 21.8 | 2.2e+17 |
| Mid | 0.0008 | 0.04 | 0.0002 | 0.99 | 0.52 | 2.6e+12 |
| Dec3 | 0.0040 | 0.09 | -0.0008 | 2.88 | 9.48 | 1.1e+12 |
| Dec2 | 0.0086 | 0.11 | -0.0019 | 4.58 | 7.52 | 8.3e+09 |
| Dec1 | 0.0193 | 0.12 | 0.0010 | 3.09 | -17.01 | 4.3e+07 |
| Dec0 | 0.0195 | 0.07 | -0.0002 | 2.13 | 8.37 | 1.2e+12 |

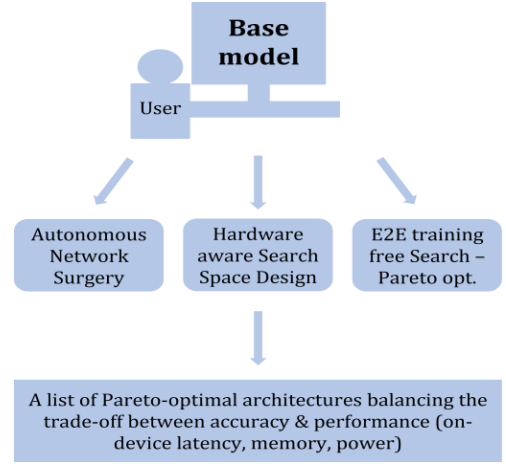

Figure 2: Flowsheet of the Proposed Method

Our optimized variants achieve comparable accuracy, up to 1.25× latency improvement on commercial smartphones with Qualcomm NPUs. Together, these results establish a principled path toward hardware-aware deblurring resulting in truly efficient edge deployment. The main contributions of this work are as follows:

1. **Framework:** A black-box, hardware-aware method that adapts restoration networks for NPUs using direct device feedback.

2. **Strategy:** A scalable approach based on sensitivity-guided network surgery and surrogate blocks distilled at the feature level.

3. **Search:** A training-free, multi-objective formulation that considers surrogate accuracy and profiled latency, solved via Bayesian optimization.

4. **Principle:** decoupling of accuracy from inference optimization, overcoming limitations of Neural Architecture Search for large networks.

5. **Validation:** Extensive experiments on motion and defocus deblurring.

## II. PROPOSED METHOD

### A. Network Surgery

The proposed method (see Fig 2) begins with a pre-trained state-of-the-art model that achieves high accuracy on the target.

Table 2. Latencies of the hardware aware alternatives on GS24

| S.No | Alt-0 | Alt-1 | Alt-2 | Alt-3 | Alt-4 | Alt-5 | Alt-6 |
|---|---|---|---|---|---|---|---|
| Latency (ms) | 53 | 15 | 13 | 19 | 11 | 9 | 28 |

Table 3. Comparison of NAFNet with Optimized model

| Model | Params | Size | Latency | PSNR | SSIM |
|---|---|---|---|---|---|
| NAFNet | 17.112 M | 107 MB | 177 ms | 35.80 | 0.980 |
| Ours | 17.110 M | 106 MB | 147 ms | 33.76 | 0.942 |

The goal is not to retrain from scratch but to restructure the model for efficient inference on embedded devices. To this end, we first convert and deploy the baseline model on the device, profiling the latency of each block. In U-Net styled architectures such as NAFNet, blocks naturally correspond to encoder, decoder, or middle modules, and their latency characteristics can be measured individually.

Alongside profiling, we estimate the saliency of each block to capture its sensitivity to accuracy. Multiple saliency [7] measures are considered, including gradient norm, SNIP, GraSP, Fisher information, plain gradient and Synflow. Blocks that exhibit high saliency are marked as critical and left unchanged, while blocks with low saliency are deemed less influential on overall accuracy.

The key idea of network surgery is then to replace these less sensitive but latency-heavy blocks with hardware-friendly alternatives. Such alternatives are drawn from prior literature and device-specific profiling. Once the candidate replacements are defined, they form the pool of surgical options to systematically substitute the troublesome blocks, enabling deployment-optimized versions of the original network.

### B. Hardware-aware Search Space Design

Following network surgery, we obtain a list of hardware-friendly alternatives that can potentially replace latency-heavy blocks. As simply substituting these blocks risks significant accuracy degradation, each alternative is converted into a digital twin of the original block, which we refer to as a *surrogate*. The purpose of these surrogates is to approximate the functional behavior of the base block while being more amenable to the target hardware.

To train these surrogates, feature-wise knowledge distillation is used. Since this procedure is independent for each block, multiple surrogates can be trained in parallel, reducing overall training time. Furthermore, because distillation is confined to block-level features rather than end-to-end outputs, the data requirement is reduced to roughly 20–25% of the full training set, further improving efficiency. To address the accuracy drop when the surrogates are stitched to become a full network, end-to-end finetuning of final model is performed.

### C. End-to-End Training-free Search

The third stage formulates the problem of selecting hardware-friendly surrogates as a multi-objective search. For a network with m blocks and n alternatives per block, the search space in principle scales as $n^m$, which is combinatorial. However, because all candidate surrogates have already been pre-trained through hardware-aware distillation, the evaluation of any configuration reduces to a plug-and-play forward pass. Specifically, replacing a block with one of its surrogates allows the immediate measurement of accuracy drop (in terms of PSNR loss) without retraining. This is the first objective.

The second objective is latency, which is estimated by profiling each block and alternative directly on the target device. Global latency is approximated as the sum of block-level latencies. To balance the scale of the two objectives, we normalize the latency into a penalty score such that its order of magnitude aligns with the accuracy loss. Thus, high-latency blocks are associated with higher penalties and low-latency blocks with smaller penalties.

With both objectives defined, we pose the problem as a multi-objective Bayesian optimization. Each network configuration is encoded as a real-valued vector, where every dimension corresponds to a block, and its value is mapped to one of the available alternatives via equal interval partitioning. The optimizer evaluates the accuracy loss and latency penalty of sampled configurations, builds Gaussian process surrogates, and uses expected hypervolume improvement to select promising candidates. This process iteratively produces a Pareto front of architectures, trading off accuracy and latency.

Finally, architectures lying in the knee region of the Pareto front are selected and fine-tuned end-to-end with full data. The best-performing model after fine-tuning is chosen as the optimized architecture. We note that, while any search strategy can be applied, the novelty of our approach lies in: (i) reformulating the combinatorial search as a training-free, plug-and-play evaluation of block surrogates, and (ii) designing the vector encoding and penalty formulation that make optimization scalable to large restoration networks.

Table 4. The generalization ability of proposed model for Image De-blurring trained on GoPro and its comparison with SOTA.

| Method | GoPro | | Real Blur-J | | Real Blur-R | | HIDE | |
|---|---|---|---|---|---|---|---|---|
| | PSNR | SSIM | PSNR | SSIM | PSNR | SSIM | PSNR | SSIM |
| **SRN** [17] | 30.26 | 0.934 | 28.56 | 0.867 | 35.66 | 0.947 | 28.36 | 0.915 |
| **DPDNet.** [12] | 29.08 | 0.914 | 27.87 | 0.827 | 32.51 | 0.841 | 25.73 | 0.874 |
| **DMPHN** [18] | 31.20 | 0.940 | 28.42 | 0.860 | 35.70 | 0.948 | 29.09 | 0.924 |
| **DBGAN** [19] | 31.10 | 0.942 | 24.93 | 0.745 | 33.78 | 0.909 | 28.94 | 0.915 |
| **MPRNet** [20] | 32.66 | 0.959 | 28.70 | 0.873 | 35.99 | 0.952 | 30.96 | 0.939 |
| **MT-RNN** [21] | 31.15 | 0.945 | 28.44 | 0.862 | 35.79 | 0.951 | 29.15 | 0.918 |
| **Restormer** [5] | 32.92 | 0.961 | 28.96 | 0.879 | 36.19 | 0.957 | 31.22 | 0.942 |
| **NAFNet** [3] | 35.80 | 0.980 | 26.48 | 0.823 | 33.78 | 0.943 | 30.45 | 0.913 |
| **Ours** | 33.76 | 0.942 | 26.51 | 0.824 | 33.82 | 0.943 | 30.36 | 0.912 |

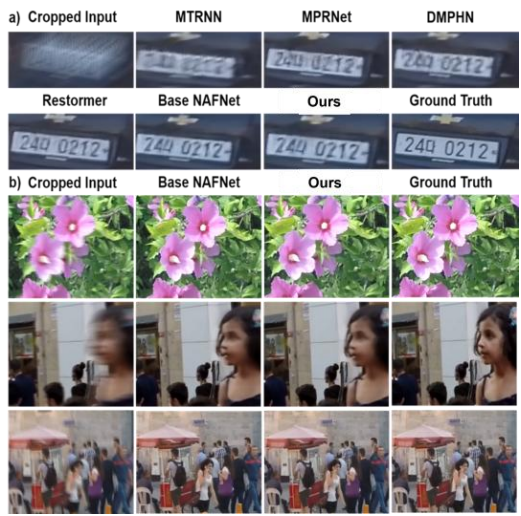

Figure 3. Visual comparison of Motion De-blur on GoPro from a) our model and SOTA and b) our and base models.

## III. EXPERIMENTS

### A. Optimization of Motion De-blurring Model

In the first set of experiments, Motion De-blurring on open source GoPro dataset [8] is considered. The pre-trained open source NAFNet [3] with 4 encoders, 1 middle block and 4 decoders is used as the base model. It consisted of 1 NAF block in each encoder, middle block and decoder, with the exception of 4th encoder which consisted of 28 NAF blocks. Excluding the sensitive 4th encoder block (confirmed by saliency scores [7]) (see Table 1), 6 different alternatives (see Fig. 1) are prepared for all other blocks in the network and profiled on GS24 Ultra (see Table 2) to assign the penalty values need during MOBO [13]. This resulted in a search space of 8 dimensions with 7 options for each dimension (6 alternatives + 1 base block) resulting in $7^8 \sim 5.7$ million combinations.

Table 5. Results on Defocus De-blurring with our model.

| Method | Indoor Scenes | | Outdoor Scenes | | MACs |
|---|---|---|---|---|---|
| | PSNR | SSIM | PSNR | SSIM | |
| **DPDNet** [12] | 26.54 | 0.816 | 22.25 | 0.682 | 991G |
| **KPAC** [22] | 27.97 | 0.852 | 22.62 | 0.701 | - |
| **IFAN** [23] | 28.11 | 0.861 | 22.76 | 0.720 | 363G |
| **DMENet** [24] | 25.50 | 0.788 | 21.43 | 0.644 | 1173 |
| **EBDB** [25] | 25.77 | 0.772 | 21.25 | 0.599 | - |
| **JNB** [26] | 26.73 | 0.828 | 21.10 | 0.608 | - |
| **Restormer** [5] | 28.87 | 0.882 | 23.24 | 0.743 | 141G |
| **KBNet** [14] | 28.89 | 0.883 | 23.32 | 0.749 | 108G |
| **NAFNet** [3] | 24.44 | 0.766 | 18.83 | 0.574 | 65G |
| **Ours** | 19.93 | 0.686 | 18.89 | 0.567 | 63G |
| **Ours** (Fine-tuned) | 26.87 | 0.818 | 21.89 | 0.656 | |

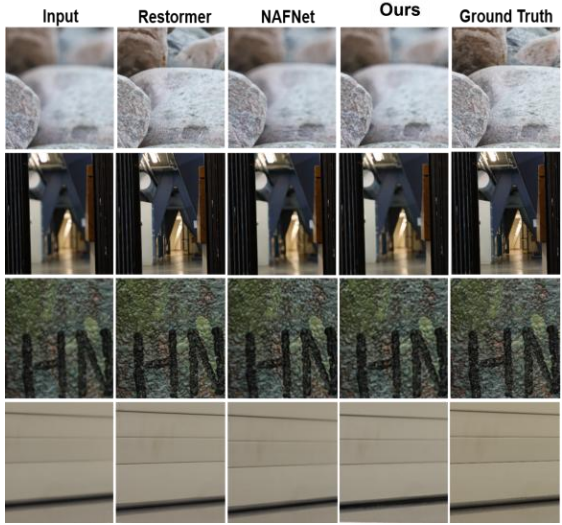

Figure 4. Visual comparison of Defocus De-blur on DPDD.

From the Pareto set obtained from MOBO [13], the network with least latency is selected as the optimized model for further analysis (see Table 3). While the base NAFNet took 177ms on GS24 Ultra, the optimized model ran in 140ms resulting in 1.25 fold improvement in on-device latency. The accuracy of the optimized model compared with the benchmarks in literature is presented in Table 4. Some visual comparison results are presented in Fig. 3.

### B. Generalization of the optimized model

The generalization ability of the optimized model can be seen in Table 4 where it is tested on the 3 datasets of motion de-blur unseen during optimization. Further, compared to the Restormer model [5] that required 140 GMACs [12, 13], the proposed network obtained better PSNR for GoPro with 68 GMACs resulting in ~55% improvement.

### C. Testing on Defocus De-blurring

The obtained model is fine-tuned for defocus de-blurring on the DPDD dataset [9] to check its adaptation to new use-case on the mobile ISP. The comparison of accuracy with SOTA in literature is presented in Table 5. The visual inspection results are presented in Fig. 4.

Compared to another recent simple baseline KBNet with 108 GMACs [14], the proposed model with 63 GMACs, has competitive accuracy. It is to be noted that while KBNet in Table 5 is trained on DPDD & the proposed model is trained on GoPro.

### D. Discussion

This work underscores the importance of hardware feedback in bridging the gap between algorithmic efficiency and deployment performance. By incorporating profiling directly into the adaptation loop, the framework avoids reliance on FLOPs or parameter counts, which often fail to correlate with on-device latency. The result is a scalable approach that produces variants aligned with actual NPU execution.

A key implication is that feedback-driven adaptation offers a practical mechanism for rapidly reconfiguring networks across devices. From an academic perspective, decoupling accuracy optimization from inference optimization suggests new avenues for studying multi-objective search under black-box hardware constraints. From an industrial standpoint, the framework provides actionable insights on module-level bottlenecks and latency trade-offs, making adaptation a repeatable process whenever new hardware or tasks emerge.

## IV. CONCLUSION

We presented a hardware-aware framework for adapting state-of-the-art image deblurring networks to mobile NPUs. The method combines sensitivity-guided network surgery, surrogate block distillation, and training-free multi-objective search with direct hardware feedback. The framework achieves comparable or improved accuracy and up to **1.25× latency improvement** on commercial smartphones. These results highlight the promise of feedback-driven strategies for producing deployment-ready restoration models that balance accuracy, efficiency, and hardware compatibility.